\documentclass[%
 reprint,
 amsmath,amssymb,
pra,
]{revtex4-2}

\usepackage{graphicx}
\usepackage{dcolumn}
\usepackage{bm}
\usepackage[mathlines]{lineno}

\usepackage{float}
\usepackage{caption}
\usepackage{subcaption}
\usepackage{upgreek}
\usepackage[english]{babel}
\usepackage{amsmath}
\usepackage{mathtools}
\usepackage{xcolor}
\usepackage{comment}

\begin{document}

\preprint{APS/123-QED}

\title{SOI-Compatible Degenerate Band Edge Photonic Structure: Design Rules and Robustness Analysis}

\author{Kessem~Zamir-Abramovich}
 \email{kessemzamir@mail.tau.ac.il}
 \author{Jacob~Scheuer}
\email{kobys@tauex.tau.ac.il}
 \affiliation{School of Electrical and Computer Engineering Tel-Aviv University, Tel-Aviv, Israel}
 \affiliation{Tel-Aviv University Center for Nanoscience and Nanotechnology, Tel-Aviv, Israel}

\date{\today}

\begin{abstract}
Optical periodic structures exhibiting a degenerate band edge (DBE) are of significant interest for various applications such as switching, sensing, high-power amplification, and lasing. At the edge of the bandgap in such structures, a fourth-order exceptional point degeneracy arises, leading to an extremely flat dispersion band. We propose and study a Silicon-on-Insulator-compatible structure composed of two coupled waveguides with asymmetric gratings. The dispersion relations and the field profiles are obtained using three-dimensional finite-difference time-domain simulations, and we provide a set of practical guidelines for the design and optimization of such structures, in order to obtain a DBE. We analyze the transmission and reflection spectra of finite-size devices, and investigate their spectral properties near the stationary points. The scaling of the resonance quality factor with the number of unit cells is studied, revealing a performance that surpasses that of conventional periodic structures. Finally, we examine the robustness of the DBE with respect to fabrication tolerances and structural imperfections.       
\end{abstract}

\maketitle

\section{\label{sec:Inrto}Introduction}

Stationary points (SPs) in periodic structures, also known as exceptional points degeneracies, correspond to frequencies (and wavenumbers) at which the group velocity approaches zero. At such points, two or more waveguide modes coalesce (both eigenvalues and eigenvectors), leading to amplification of various properties of the photonic structure such as transmission, gain, Q factor and absorption \cite{SL_absorption2007, low_vg__effects_Kiyota2006, ultraslow_group_velocity_1999,Band_edge_laser_Dowling1994,anisotropic_layers_Figotin2005,Periodic_Coupled_WG_Abdelshafy2019,zamir2023lowthreshold,hasanli2025ep_passive_strip_waveguides}. 
The order of an SP is defined according to the number of eigenmodes that coalesce. The simplest SP is of order 2, which is formed, e.g., at the band-edge of Bragg gratings. Such SP is also called a regular band edge (RBE) SP. SPs with higher orders exist as well. Particularly, $4^{th}$ order SPs, also known as degenerate band edge (DBE) SPs, have drawn significant interest in the past decade due to their potential applications for switching and sensing, as well as  high-gain amplifiers and low-threshold, narrow-linewidth, lasers ~\cite{brimont2012slowlight_modulator,wang2010slowlight_sensing,veysi2018band_edge_laser}. 
One of the signature properties of SPs is the dispersion relations in their vicinity, which exhibit a polynomial profile of order $M$, where $M$ is the order of the SP. For a DBE SP, the dispersion relations are characterized by a quartic profile, $\omega-\omega_D\propto(k-k_D )^4$, where $\omega_D$ and $k_D$ are, respectively, the frequency and wavenumber of the SP. The DBE is part of a wide phenomenon called the frozen mode regime, characterized by the coalescence of both propagating and evanescent modes of the structure. As opposed to an RBE, in high order exceptional points such as the DBE, the amplification of properties is due to contribution of the evanescent waves as well as propagating waves  \cite{figotin2011slow_wave_phenomena,Capolino2017theory_resonators}.
In the past two decades, several periodic photonic structures exhibiting DBEs, have been proposed and studied: subwavelength resonators \cite{kornovan2021high_Q_localized}, coupled ring resonators \cite{Capolino2018anomulousQ}, coupled waveguides with cylindrical holes \cite{burr2013Si_WG_with_holes}, coupled waveguides with gratings \cite{mealy2022distributed_feedback_gratings} and waveguide ridge with etched holes \cite{wood2015ridge_WG_with_holes}. A general analysis of the dispersion relation of periodic structures and the conditions for obtaining SPs (including DBEs) was carried out in \cite{gutman2012slowlight_gratings_analytical}. An analysis of the dispersion relations of specific photonic structures exhibiting DBEs was carried numerically, mainly for a two-dimensional case \cite{sukhorukov20082D_WGs_with_holes_numerically,mealy2022distributed_feedback_gratings}. Analysis of three-dimensional structures was done only scarcely \cite{burr2013Si_WG_with_holes, wood2015ridge_WG_with_holes}. However, the robustness of those structures to geometrical imperfections was not considered.
\begin{figure*}[btp]
    \centering
    \includegraphics[scale=0.7]{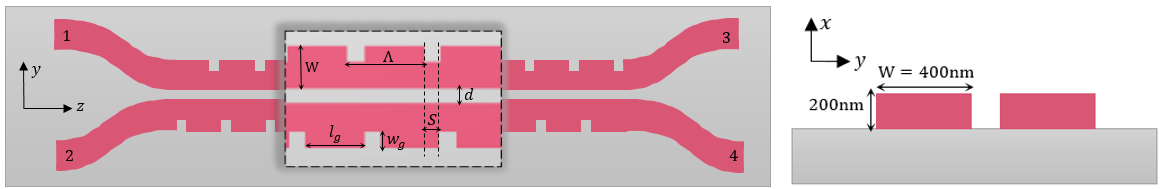}
    \caption{Schematic of the proposed structure, composed to two coupled Silicon waveguides with gratings, on SiO$_2$ substrate.}
    \label{fig:structure}
\end{figure*}

While the fundamental benefits of operating at a DBE SP are clear, designing such a structure is not trivial and often requires the determination of several design parameters, necessitating numerous trial-and-error optimization steps. Furthermore, the necessity to determine the specific values of the design parameters which lead to the formation of a DBE raises a major concern regarding the robustness of design (and the DBE) to errors and fabrication tolerances. This is in contrast to RBEs, that are inherently robust and are necessarily formed in any periodic structure that couples forward and backward propagating waves.

In this paper, we present and analyze a realistic, standard Silicon-on-Insulator (SOI) process compatible , photonic structure consisting of two coupled waveguides having asymmetric gratings (see Fig.~\ref{fig:structure}). We present a set of guidelines facilitating the design of a photonic structure exhibiting a DBE SP and use them to present structure employing both weak and strong coupling between the waveguides. We study the dispersion relations of these structures near DBEs, showing the quartic relation between the frequency and the wavenumber. We also study the transmission and reflection properties of finite structures employing our design. We identify the resonances associated with the DBE and show that their Q-factor scales as $N^5$, clearly indicating the formation of a DBE (see Section \ref{sec:Finite_structure} for more details). Finally, we study the robustness of the dispersion relations near the DBE to errors and biases in the design parameters. We show that they are robust to reasonable fabrication tolerances of contemporary nano-fabrication techniques.

The rest of the paper is organized as follows. In section \ref{sec:Disp} we present a strategy for optimizing the structure’s geometry to obtain a DBE SP, and present the resulting dispersion relations. In section \ref{sec:Finite_structure}  we analyze the spectral transmission and reflection of structures with finite number of unit cells. In Section \ref{sec:Tolerence} we study the robustness of the structure to fabrication errors and in section \ref{sec:conclusion} we summarize and conclude.

 \section{\label{sec:Disp}Dispersion Relations}
The proposed periodic structure consists of two coupled waveguides with periodic corrugations (see Fig. ~\ref{fig:structure}). The structure is based on standard Silicon on Insulator (SOI) layers, where the waveguides employ a 400nm by 200nm Si core placed on an SiO$_2$ lower clad. The spacing between the waveguides is designated as $d$, the gratings are rectangular with length of $l_g$ and width of $w_g$. The periodicity of the gratings is $\Lambda$. It should be noted that the gratings reduce, locally, the width of the waveguides. The gratings on each waveguide are not a mirror image of each other, but rather exhibit an offset in the $z$ direction denoted as $S$. This offset is crucial for obtaining a DBE, because it allows for coupling between the even and odd modes of the structure, as discussed below. Each waveguide supports a single TE mode.  The dispersion relations and Bloch wave solutions (as well as all other simulations of the structure) were calculated using Ansys Lumerical$^{\mathrm{TM}}$ FDTD tool. 
 \begin{figure}[h]
    \centering
    \begin{subfigure}[b]{0.4\textwidth}
    \centering
     \includegraphics[scale=0.6]{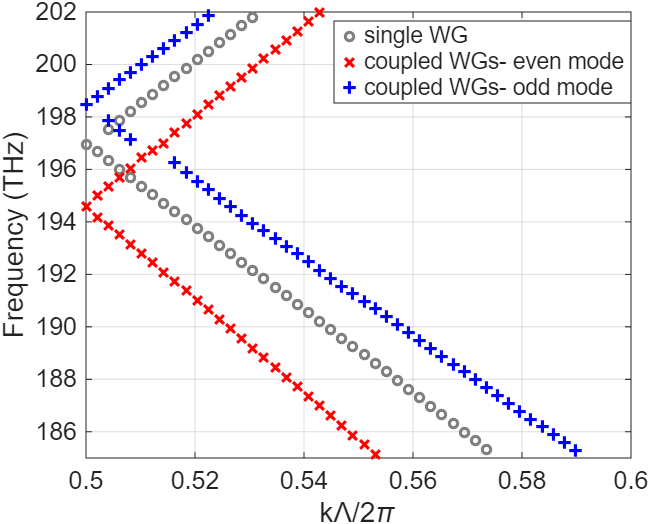}
    \subcaption{}    \label{fig:two_coupled_WG_vs_one_WG}
    \end{subfigure}
    \hfill
    \begin{subfigure}[b]{0.4\textwidth}
    \centering
     \includegraphics[scale=0.6]{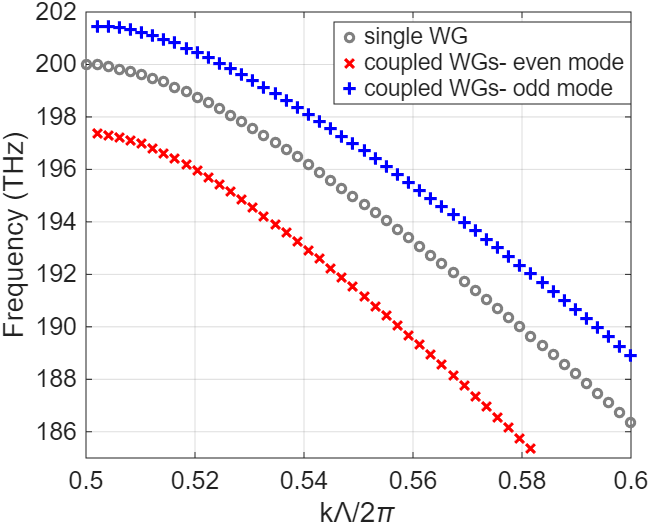}
    \subcaption{}    \label{fig:two_coupled_WG_gratings_vs_one_WG_gratings}
    \end{subfigure}
    \caption{Dispersion relations of a single waveguide (gray circles) and two coupled waveguides (red `x' - even mode, blue `+' - odd mode). (a) Waveguides without gratings. (b) Waveguides with gratings.}
    \label{fig:dispersions}
\end{figure}

DBEs are formed when four eigenvectors and their corresponding eigenvalues coalesce. For that to occur, power must be exchanged efficiently between the four modes. The geometrical parameters of the structure depicted in Fig.~\ref{fig:structure} must be designed to support such power exchange. As the design space consists of 5 different parameters; it is useful to present briefly a design flow that can leads to a formation of a DBE in the Bloch dispersion relations. To start, consider first the eigenmodes of two coupled single-mode waveguides (no gratings). In this simple case, the structure supports two orthogonal TE modes, exhibiting even and odd profiles. Each mode can propagate either forward or backward, thus yielding a complete orthogonal basis consisting of four waves. We mark the propagating coefficients of the even and odd modes as $\beta_e$ and $\beta_o$, respectively. Using this basis simplifies the structure optimization because the coupling between the two waveguides (related to the distance between them, $d$) is embodied in the propagation coefficients. The dispersion relations of the two coupled single mode waveguides and that of a single mode waveguide is shown in Fig.~\ref{fig:two_coupled_WG_vs_one_WG}. The red `x' and blue `+' correspond to the dispersion of the even and odd modes of two coupled waveguides, respectively. The gray circles indicate the dispersion relations of a single mode waveguide. As can be expected, the coupling induces splitting of the dispersion curve into two distinct modes (even and odd). Note that the splitting is not symmetric around the single waveguide dispersion curve. The asymmetry arises from the self- frequency shift \cite{yariv2007photonics}. Thus, we can express $\beta_e$ and $\beta_o$ as a combination of a symmetric splitting of single waveguide mode propagation coefficient, $\beta_s$, due to the coupling and shift caused by the self-frequency shift:
  \begin{equation}\label{eq:1}
      \beta_{e,o}=\beta_s\pm\frac{1}{2}\Delta\beta_{\kappa}+\frac{1}{2}\Delta\beta_{\delta}
 \end{equation}

Next, we introduce gratings to the structure to couple the forward and backward propagating waves, but without offset. The structure is similar to that of Fig.~\ref{fig:structure}, but with $S=0$. In this configuration, the coupling between the even and odd modes is inefficient. However, the coupling between forward and backwards propagating modes of the same parity can be made efficient if the period of the gratings satisfies the relations in Eqs.~\ref{eq:2},\ref{eq:3}. We note that the period of the gratings also defines the period of the unit cell $\Lambda$. Fig.~\ref{fig:two_coupled_WG_gratings_vs_one_WG_gratings} shows the dispersion relation of two coupled waveguides with mirrored gratings. Similar to  Fig.~\ref{fig:two_coupled_WG_vs_one_WG}, the dispersion relations of the even and odd modes is indicated by red `x' and blue `+', respectively. The gray circles indicate the dispersion curve of a single waveguide with gratins of the same period and dimensions. As for the grating-less case, the dispersion curve splits into two because of the coupling, and RBEs are formed at $k\Lambda=\pi$. Note that because $S=0$ here, coupling can be attained only between forward and backward propagating modes of the same parity. Hence, coalescence of only two modes can be achieved, resulting in the formation of RBEs. 

The final step is the introduction of an offset between the gratings of the two waveguides. Due to this offset, the even and odd modes can couple to each other, thus providing the necessary conditions for the formation of a DBE. To obtain efficient coupling between modes with different parity, it is not sufficient to break mirror symmetry, it is also necessary to choose the periodicity to obtain phase matching between the modes. This can be attained when Eq.~\ref{eq:4} is satisfied. Note that the condition of Eq.~\ref{eq:4} ensures coupling between \textit{counterpropagating} modes of different parity. Although it is possible to couple the co-propagating modes of the different parity by choosing larger period (replacing the plus sign in the LHS of Eq.~\ref{eq:4} with a minus sign), this would eliminate the coupling between counterpropagating waves which is also necessary for forming a DBE. Eqs.~\ref{eq:2}-\ref{eq:4} provides a framework for determining the range of the periodicities that can facilitate the formation of a DBE. However, for that purpose, the gratings geometry and $S$ must also be optimized.
\begin{align}
     2\beta_e\approx \frac{2\pi}{\Lambda}\label{eq:2}\\
    2\beta_o\approx \frac{2\pi}{\Lambda}\label{eq:3} \\
    \beta_e+\beta_o\approx \frac{2\pi}{\Lambda}\label{eq:4}
\end{align}

   \begin{figure}[h]
    \centering
    \begin{subfigure}[b]{0.23\textwidth}
    \centering
     \includegraphics[scale=0.4]{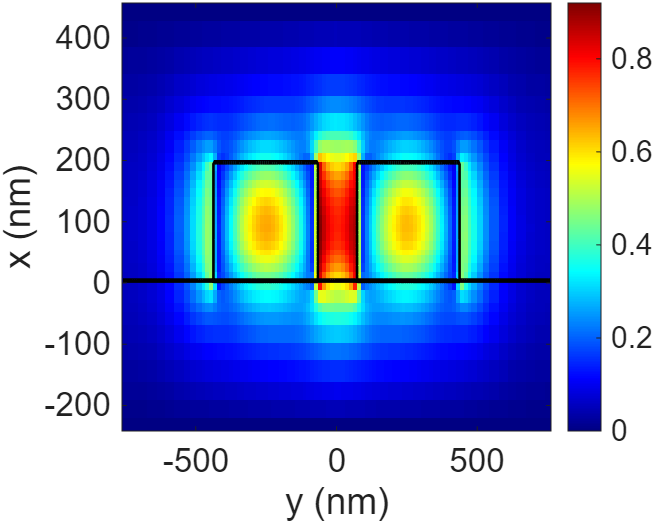}
    \subcaption{}
    \label{fig:even_mode}
    \end{subfigure}
    \hfill
    \begin{subfigure}[b]{0.23\textwidth}
    \centering
     \includegraphics[scale=0.4]{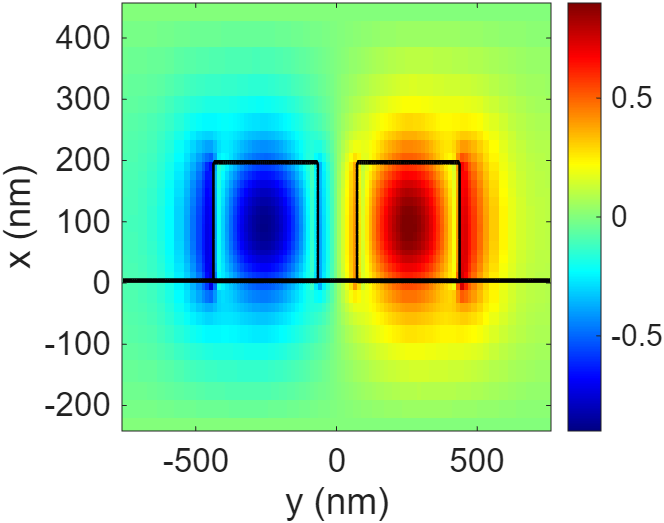}
    \subcaption{}
    \label{fig:odd_mode}
    \end{subfigure}
    \caption{TE eigen-modes of two coupled waveguides. The panels show a cross section of real part of $E_y$ in the xy plane. (a) Even mode. (b) Odd mode.}
    \label{fig:even_odd_mode}
\end{figure}
   
Choosing $\Lambda$ such that Eqs.~\ref{eq:2}-\ref{eq:4} are satisfied, requires the knowledge of the propagation coefficients of the even and odd modes, which are determined by the waveguides dimensions and spacing. The first step is to choose the spacing between the waveguides, $d$. Although this choice is somewhat arbitrary, it is advantageous to choose a realizable spacing (i.e. not too small). However, as further discussed below, there are some benefits for obtaining strong coupling between the waveguides, a consideration which sets an upper limit on $d$. The next step is to choose the effective width, $W_{eff}$, of the waveguide. $W_{eff}$ represents the mean value of the waveguide with the gratings. Referring to Fig.~\ref{fig:structure}, the gratings effectively reduce the width of the coupled waveguides according to their width and duty-cycle, which means that $W_{eff}<W$. With the choice of $d$ and $W_{eff}$, one can obtain $\beta_e$ and $\beta_o$ which, in turn, allows for choosing $\Lambda$. The third step is choosing the parameters of the gratins, $w_g$ and $l_g$. One of the constraints on these parameters is the choice of $W_{eff}$. As the periodicity of the gratings is of subwavelength dimensions (approximately half-wavelength in the material), a rough approximation for the effective width can be obtained by:
\begin{equation}\label{eq:5}
    W_{eff}=\frac{w_g}{\Lambda}\cdot W+\frac{\Lambda-w_g}{\Lambda}\cdot (W-h_g)
\end{equation}

 \begin{table}
    \caption{Geometrical parameters for structures exhibiting DBEs. Set 1 - weak coupling between modes, Set 2 - strong coupling between modes}
\begin{ruledtabular}
    \begin{tabular}{ccccccc}
        Parameter (nm) & $\Lambda$ & $d$ & $w_g$ & $l_g$ & $S$ & $W$\\
        \hline
        Set 1 & 395 & 300 & 82 & 215 & 135 & 400\\
        Set 2 & 400 & 150 & 150 & 300 & 100 & 400\\            
    \end{tabular}
    \label{table: geometrical_perameters}
\end{ruledtabular}
\end{table}

\begin{figure}[h]
    \centering
   \begin{subfigure}[b]{0.22\textwidth}
    \centering
     \includegraphics[scale=0.75]{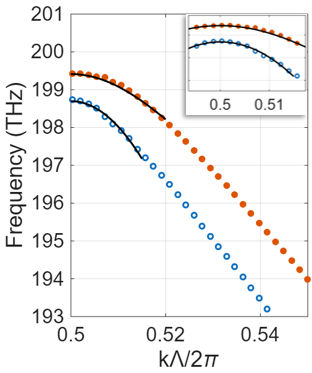}
    \subcaption{}    \label{fig:weak_disp}
   \end{subfigure}
    \hfill
   \begin{subfigure}[b]{0.22\textwidth}
    \centering
     \includegraphics[scale=0.75]{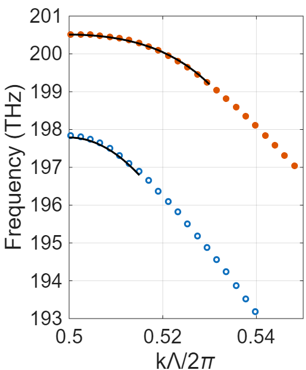}
    \subcaption{}    \label{fig:strong_disp}
   \end{subfigure}
    \caption{Bloch dispersion relations corresponding to Set 1 and Set 2 designs. Orange dots and light blue circles mark a DBE and an RBE at the center of the Brillouin zone, respectively. (a) Weak coupling case (Set 1). inset - zoom in on the DBE and RBE at the center of the Brillouin zone. (b) Strong coupling case (Set 2).}
    \label{fig:weak_strong_disp}
\end{figure} 

Eq.~\ref{eq:5} yields only a constraint on $w_g$, $l_g$, and $\Lambda$, thus allowing for an infinite number of solutions. Another useful rule of thumb that can be used here is the strength coupling between parallel waveguides. Generally, the strength of the backward scattering (caused by the gratings) should be of the same order of magnitude as that of the coupling between the waveguides (determined by $d$). In other words, for strong coupling between the waveguides, gratings exhibiting stronger perturbations should be chosen (see also table~\ref{table: geometrical_perameters} below). The last step is to optimize the offset between the waveguides $S$, in order to obtain a DBE in the dispersion relations.

As a concrete example, we consider standard SOI layer structure (see Fig.~\ref{fig:structure}). To ensure material transparency we choose to work at the telecom band with $\lambda\sim 1.5 \mu\textrm{m}$. We choose an effective waveguide of $W_{eff}=362.5\textrm{nm}$ and consider two values of waveguides separation: $d=150\textrm{nm}$ (strong coupling) and $d=300\textrm{nm}$ (weak coupling). Table~\ref{table: geometrical_perameters} lists the corresponding sets of design parameters found following the process described above. Set 1 (weak coupling) yields effective indices of $n_e=1.91$ and $n_o=1.89$, while set 2 (strong coupling) yields $n_e=1.96$ and $n_o=1.90$. Using these these values and using Eq.~\ref{eq:1} we obtain $\Delta\beta_{\kappa}=0.087 \textrm{rad/}\mu \textrm{m}$ and $\Delta\beta_{\kappa}=0.289\textrm{rad/}\mu \textrm{m}$ for set 1 and 2 respectively. As can be expected, the wavenumbers separation is significantly larger for set 2 (strong coupling).
Fig.~\ref{fig:even_odd_mode} shows the real part of $E_y$ of the even and odd modes of two coupled waveguides with $W_{eff}=362.5\textrm{nm}$ and $d=150\textrm{nm}$ (i.e. set 2). It should be noted that the field amplitude is maximal in the \emph{space} between the two waveguides. This is a clear indication for strong coupling between the waveguides, a scenario which is not described accurately by coupled mode theory. Having found $n_e$ and $n_o$, we continue with the procedure described above to set the $\Lambda$, the gratings parameters $l_g$ and $w_g$, and the offset parameter $S$.
Fig.~\ref{fig:weak_disp} and \ref{fig:strong_disp} depict the Bloch dispersion relations found for sets 1 and 2, respectively. In both cases, two bands are observed, where the higher (lower) band exhibits a DBE (RBE) at $k\Lambda=\pi$. The bands exhibiting DBEs and RBEs are marked by orange and blue circles, respectively. Specifically, for set 1 (Fig.~\ref{fig:weak_disp}) the DBE is obtained at $\thicksim 199.5\textrm{THz}$ ($\thicksim 1504 \textrm{nm}$), and in Fig.~\ref{fig:strong_disp} the DBE is at $\thicksim 200.5 \textrm{THz}$ ($\thicksim 1496\textrm{nm}$). To verify that the obtained bands indeed exhibit a DBE and an RBE, we fitted the dispersion relations in the vicinity of $k\Lambda=\pi$ to a $4^{th}$ order polynomial function for both the upper and lower bands (orange and blue circles respectively) in Fig.~\ref{fig:weak_strong_disp} according to Eq.~\ref{eq:6}:

\begin{equation}\label{eq:6}\\
     \omega-\omega_{SP}=a_{SP}\cdot(\hat{k}-0.5)^4+b_{SP}\cdot(\hat{k}-0.5)^2 
\end{equation}  

where $\hat{k} = {k}\frac{\Lambda}{2\pi}$ is the normalized wavenumber. The subscript $SP$ corresponds to the type of the stationary point and can be either $RBE$ or $DBE$. The fitting parameters for the dispersion relation for each set in Fig.~\ref{fig:weak_strong_disp} are detailed in Table ~\ref{tab:fitting_parameters}. For the upper band fit, near the DBE frequency $a_{DBE}$ is the dominant coefficient for both sets. This coefficient is larger by two  orders of magnitude than the second order term in Eq.~\ref{eq:6}. On the other hand, for the lower band fit $b_{RBE}$ is the dominant coefficient, and $a_{RBE}$ is zero in both cases. Clearly, the higher frequency bands exhibit quartic profiles and the lower frequency bands exhibit parabolic profiles, thus indicating the formation of DBEs and RBEs at $k\Lambda=\pi$, respectively.
Referring to Fig.~\ref{fig:weak_disp}, note that the gap between the DBE and RBE is smaller compared to Fig.~\ref{fig:strong_disp}. This is because of the stronger coupling between the waveguides in the structure corresponding to set 2. Consequently, in a practical, finite, structure the resonances corresponding to the RBE and the DBE frequencies would be further separated. In addition, note that in the weak coupling case, the momentum range where the DBE band is flat, is smaller compared to that obtained with strong coupling. As a result, a device based on set 1 must be longer than a device based on set 2 in order to exhibit DBE properties. Hence, we deduce that for practical applications, working at strong coupling conditions is advantageous.
 
\begin{table}
    \caption{Fitting parameters for Eq.~\ref{eq:6} corresponding to the dispersion relations in Fig.~\ref{fig:weak_strong_disp}}
\begin{ruledtabular}
    \begin{tabular}{ccccc}
        Parameter (nm) & $a_{DBE}$ & $b_{DBE}$ & $a_{RBE}$ & $b_{RBE}$\\
        \hline
        Set 1 & $-1\cdot 10^6$ & $-3\cdot10^4$ & $0$ & $-7\cdot 10^4$\\
        Set 2 & $-6\cdot10^5$ & $-1\cdot10^3$ & $0$ & $-4\cdot10^3$\\         
        
    \end{tabular}
    \label{tab:fitting_parameters}
\end{ruledtabular}
\end{table}

\begin{figure}[h]
    \centering
   \begin{subfigure}[b]{0.4\textwidth}
    \centering
     \includegraphics[scale=0.6]{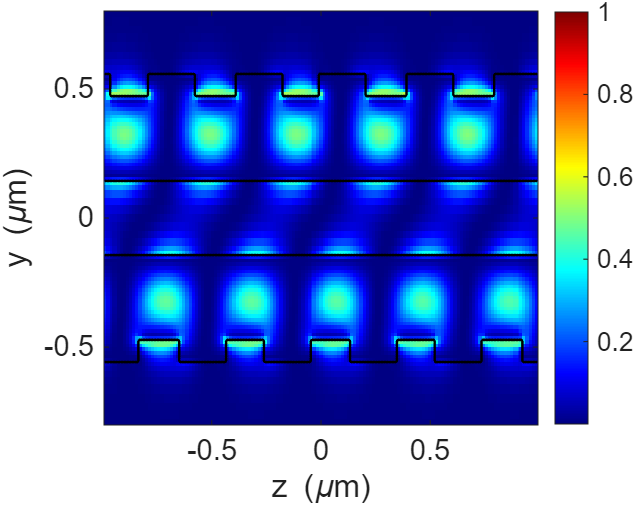}
    \subcaption{}    \label{fig:weak_field}
   \end{subfigure}
    \hfill
   \begin{subfigure}[b]{0.4\textwidth}
    \centering
     \includegraphics[scale=0.6]{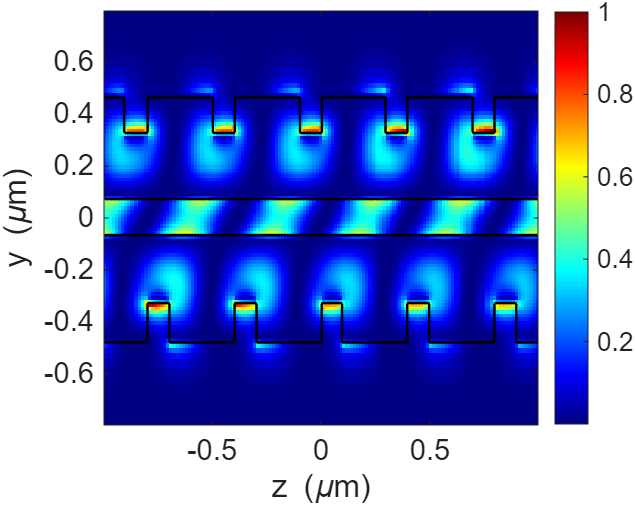}
    \subcaption{}    \label{fig:strong_field}
   \end{subfigure}
    \caption{{Electrical field intensity, $|E_y|^2$, profile inside the structure, near the DBE frequencies (a) Geometrical parameters of Set 1. (b) Geometrical parameters of Set 2.}}
    \label{fig:weak_strong_field}
\end{figure} 

Fig.~\ref{fig:weak_strong_field} depicts the profiles of $|E_y|^2$ inside the structure, close to the DBE’s frequency, for each set (see caption for details). The profiles are normalized in both cases such that they carry the same power. For the weak coupling case, the field is concentrated mainly in the core of the two waveguides. In contrast, for the strong coupling case, the field is maximal in the gap between the waveguides (as can be expected from Fig.~\ref{fig:even_mode}) and near the edge of the gratings. In addition, the intensity near the edge of the gratings is as twice as large for the strong coupling case than for weak coupling. The latter is caused by the relatively strong perturbation induced by the gratings in set 2. In both cases, the field profiles in the waveguides exhibit a longitudinal shift. This is due to the offset between the gratings introduced to the individual waveguides.      

\section{\label{sec:Finite_structure}Finite Structure Analysis}
\begin{figure}[h]
    \centering
     \includegraphics[scale=0.5]{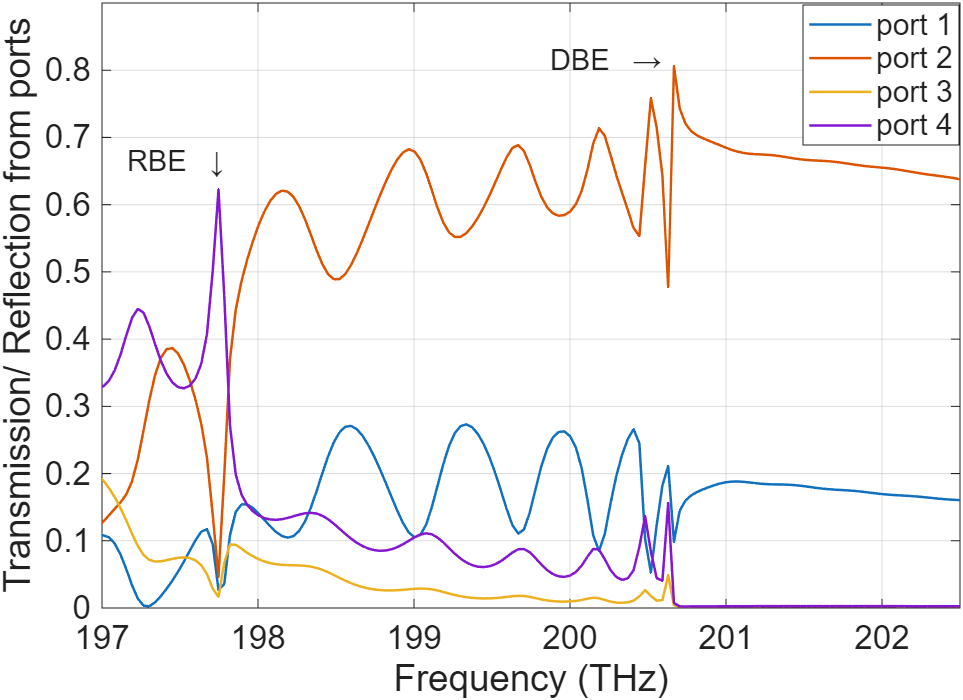} 
    \caption{Transmission and reflection from a finite structure (80 unit-cells) corresponding to Set 2.}
    \label{fig:T_R_80UC}
\end{figure}

As discussed above, the structure with strong coupling between the waveguides (and, hence, between all modes) is advantageous for practical applications. Therefore, we focus on the structure with the parameters of set 2. A realistic structure, consisting of a finite number of unit cells, includes four I/O ports as shown in Fig.~\ref{fig:structure}. In such structures, the enhancement of light near a stationary point increase with the number of unit cells, $N$. Because of the inherent impedance mismatch between the device and the I/O ports, such structures usually exhibit Fabry–Perot (FP)-like resonances. In the vicinity of stationary points, these resonances become narrower and denser due to extremely low group velocity. The longer the structure, the smaller free spectral range (FSR), leading to the formation of resonances closer to the SP. As a result, as the structure is increased, the closest resonance to the SP becomes narrower, thus possessing higher Q-factor.  For band-edge SPs (SPs of even order) this Q-factor has been shown to scale with the number of unit-cells as $Q \sim N^{M+1}$, where $M$ is the SP order \cite{Capolino2018anomulousQ, kornovan2021high_Q_localized, nada20206DBE}. Thus, by plotting the Q-factor as a function of $N$ it is possible to verify the existence of a SP, and obtain its order. 
As the structure is almost symmetrical, we consider its response to excitation from a single port. Specifically, we excite the structure from port 2 (see Fig.~\ref{fig:structure}) and examine the transmission and reflection in all four ports. Fig.~\ref{fig:T_R_80UC} shows the spectral outputs from all ports, defined as $\frac{|E_{out}|^2}{|E_{in}|^2}$,  of a structure consisting of 80 unit cells. It is important to note that, despite the large number of unit cells, the structure is only $32 \mu\textrm{m}$ long. 

\begin{figure}[h]
    \centering
     \includegraphics[scale=0.6]{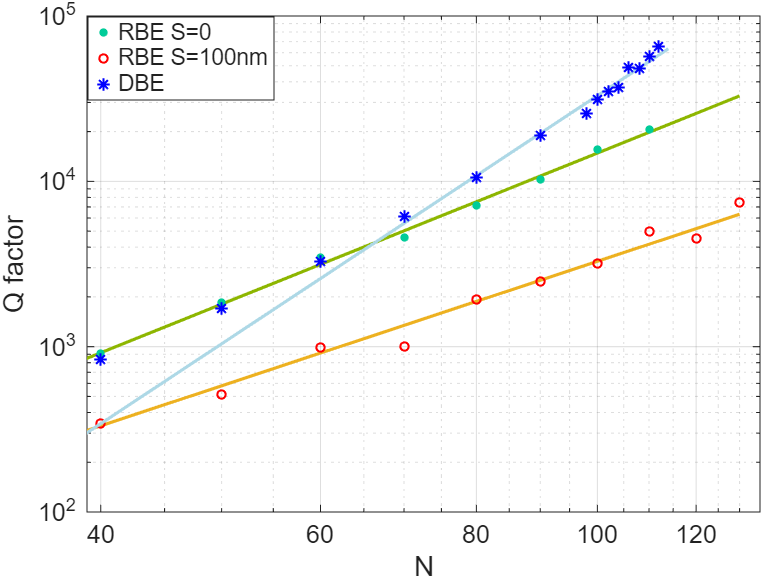} 
    \caption{{Q- factor of the DBE (blue stars) and RBE (red circles) resonances scaling with $N$ for device based on Set 2. Axes are shown in logarithmic scale. Green dots - the resonance of a RBE in a structure without offset between the gratings.}}
    \label{fig:Q_factor}
\end{figure}

\begin{figure*}[btp]
    \centering
     \includegraphics[scale=0.95]{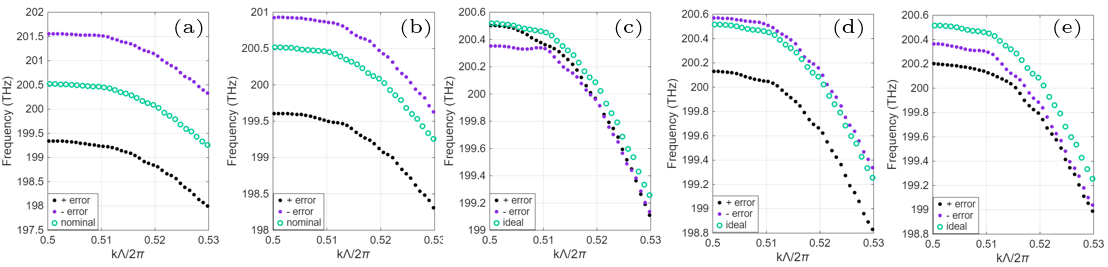}
       \caption{Change in the dispersion relation due to systematic geometrical errors. Green circles - dispersion with the nominal parameters (Set 2). Black dots increase in parameters magnitudes. Purple dots - decrease in parameters magnitudes. (a) change in the waveguides width $W$. (b) Change in $l_g$. (c) change in $d$. (d) change in $w_g$. (e) change in $S$.}
    \label{fig:errors_tolerance}
\end{figure*}

The spectra depicted in Fig.~\ref{fig:T_R_80UC} exhibit several resonances. A detailed inspection of the spectra reveals two distinct regions: below $200.7\textrm{THz}$, multiple resonant peaks, that become denser and sharper as the frequency is closer to the border frequency, are observed. Above that frequency, the output from ports 3 and 4 become negligible and no resonances are observed at the output from ports 1 and 2. The observed characteristics can be understood directly from the dispersion relations (Fig.~\ref{fig:strong_disp}). Within the bandgap ($\nu>200.7\textrm{THz}$), light cannot propagate through the structure resulting in high reflectivity and no oscillations. Below that frequency, light can propagate through the structure, forming FP-like oscillations in the transmission (and reflection) spectra due to the impedance mismatch between the Bloch modes of the device and the I/O ports. 
Next, consider the two sharp peaks at $\nu=197.7 \textrm{THz}$ (port 4) and $\nu=200.7\textrm{THz}$ (port 2). We identify these peaks as the FP resonances closest to the RBE and the DBE frequencies, respectively (see Fig.~\ref{fig:strong_disp}). Note that these resonances are narrower and exhibit higher transmission compared to the other resonances in their vicinity, owing to the very slow group velocities near stationary points. As can be expected, the resonance near the DBE SP is sharper (i.e., possessing a larger Q-factor) than that near the RBE SP. As noted above, the Q-factor of resonances in the vicinity of a band edge SP scales with the number of unit cells, as the power of the SP order plus 1. Thus, the Q-factor of the resonances near the RBE and DBE SPs should scale as $N^3$ and $N^5$, respectively. To verify that, we calculated the Q-factor of the resonances closest to the two stationary points (the RBE and DBE) for various values of $N$ for the structure with the design of set 2. Fig.~\ref{fig:Q_factor} shows the Q-factor as a function of $N$ for the RBE (red circles) and DBE (blue stars) resonances in logarithmic scale. Once $N$ exceeds a value of $\sim 60-70$, the relation between the Q-factor and $N$ shows a clear power-law characteristics. A linear fit to the calculated data points shows that Q-factors of the resonances in the vicinity of the RBE and DBE SPs increase as $\sim N^{2.5}$ and $\sim N^5$ respectively. The results are very close to the expected theoretical values, indicating the formation of an RBE in the lower band and a DBE in the upper band. It is important to note that the RBE we are referring to in Fig.~\ref{fig:strong_disp} is not a real band-edge SP, since there is no optical bandgap above it. Instead, at this frequency point we have two counter-propagating waves and one standing wave (with zero group velocity). This may be the cause for the scaling of the Q- factor with $N$, which slightly deviates from the theoretical value of $3$. To explore this further, we calculate the Q-factor as a function of the number of unit cells for a similar structure without shift between the waveguides (i.e. $S=0$) . The dispersion of such structure was shown in Fig.~\ref{fig:two_coupled_WG_gratings_vs_one_WG_gratings}, and the RBE resonance is obtained at $\nu=201.5\textrm{THz}$. The scaling of the Q- factor with $N$ is plotted in Fig.~\ref{fig:Q_factor} (green dots). A linear fit in this case yields an increase of the Q-factor that scales as $N^3$ as expected. Moreover, below $N\sim 60$, the scaling of the DBE resonance Q-factor of set 2 and the that of the RBE for $S=0$ is practically identical. Nevertheless, as the number of unit cells increases the Q-factor of the resonances associated with the DBE SP becomes greater.   

\section{\label{sec:Tolerence}Robustness to Geometrical and Fabrication Imperfections}

\begin{figure*}[btp]
    \centering
     \includegraphics[scale=1]{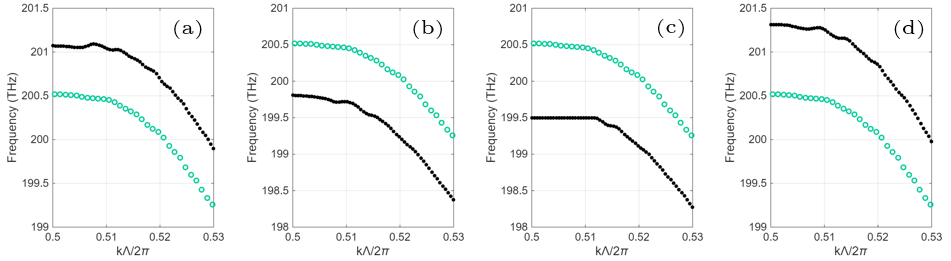}
       \caption{Change in the dispersion relation due to simultaneous geometrical errors. Green circles - the dispersion with the nominal parameters (Set 2). Black dots - perturbed parameters. (a) change: $w_g\;-2\textrm{nm}$, $l_g\; \textrm{and}\;d\;-5\textrm{nm}$, $S\;+5\textrm{nm}$. (b) Change: $w_g\;+2\textrm{nm}$, $l_g\;\textrm{and}\;d\;+5\textrm{nm}$, $S\;-5\textrm{nm}$. (c) change: $w_g\;+2\textrm{nm}$, $l_g,\;d\;\textrm{and}\;S\;+5\textrm{nm}$. (d) change: $w_g\;-2\textrm{nm}$, $l_g,\;d\;\textrm{and}\;S\;-5\textrm{nm}$.}
    \label{fig:errors_tolerance_simultaneous_change}
\end{figure*}

Any realistic periodic structure is expected to exhibit geometrical variations due to fabrication tolerances and errors. Thus, we investigate the impact of systematic geometrical errors on the dispersion curve near the DBE frequency. We consider errors of $\pm 5\textrm{nm}$ in several parameters of the structure: $W$, $d$, $l_g$ and $S$. As the gratings themselves are rather small we consider errors of $\pm2\textrm{nm}$ in $w_g$. The periodicity of the structure is defined by lithographic process which is of high accuracy. Therefore, we assume that the length of the unit cell (i.e. the periodicity) retains its nominal value.   

We start by exploring the impact of individual design parameter variation on the dispersion relations. Fig.~\ref{fig:errors_tolerance} shows the modification in the dispersion curves due to variations in a single design parameter. In each panel, the green circles indicate the dispersion relations of the nominal design (set 2). The black (purple) dots show the modified dispersion relation due to positive (negative) change of $5\textrm{nm}$ / $2\textrm{nm}$ in the relevant parameter. Focusing on the dispersion curve in the vicinity of the DBE frequency, it can be seen that the curve remains flat. An analysis of the five plots confirms that the fourth-order profile of dispersion relation is preserved under the perturbation, although the frequency of the DBE changes. It should be emphasized that the robustness of the DBE SP to perturbations is not obvious. Unlike RBEs, whose formation is guaranteed by periodicity, the formation of a DBE requires a specific set of design parameters. Nevertheless, the analysis shows that the properties of the DBE are retained even when small perturbations are introduced to the structure geometry.  
Fig.~\ref{fig:errors_tolerance}a shows the dispersion relations following a modification of the waveguides’ width. The primary impact of the perturbation is on the frequency where the DBE is obtained, while the curve profile is hardly modified. Specifically, an increase of $+5\textrm{nm}$ in $W$ red-shifts the DBE by $\sim1\textrm{THz}$, and a similar decrease in this parameter blue-shifts it by $\sim1 \textrm{THz}$.  Fig.~\ref{fig:errors_tolerance}b depicts the impact of variations in $l_g$ on the dispersion relations. As for the  gratings length, the main impact is on the DBE frequency ($\pm5\textrm{nm}$ change leads $-1 / +0.5 \textrm{THz}$ shift in the frequency). The modifications of the curve due to such perturbations are minor.  Fig.~\ref{fig:errors_tolerance}c shows the impact of variations in the waveguide spacing $d$, which corresponds to the coupling between the two waveguides. Unlike perturbations in $W$ and $l_g$, this parameter has relatively small impact on the DBE frequency, but significant impact on the dispersion curve near the DBE. The impact of variations in $w_g$ on the dispersion curve is shown in Fig.~\ref{fig:errors_tolerance}d. Recall that here, the perturbation introduced to the design parameter is only $\pm2\textrm{nm}$ because of the relatively small dimensions of $w_g$. As shown in the figure, the impact of the dispersion curve shape is minimal and a red-shift of $\sim0.4 \textrm{THz}$ in the DBE frequency is observed when $w_g$ is increased by $2\textrm{nm}$. Interestingly, minimal impact on the dispersion profile and DBE frequency is observed when $w_g$ is decreased.  Finally, Fig.~\ref{fig:errors_tolerance}e shows the impact of the grating shift parameter, $S$ on the device dispersion curve. Both positive and negative variations yield a negative spectral shift in the DBE frequency. However, a negative variation also introduces a modification of the dispersion profile manifested by a small linear slope of the curve near the DBE SP.

Now that we understand the impact of the individual parameters on the dispersion relations, let us consider a realistic scenario where variations are introduced to all the design parameters. Fig.~\ref{fig:errors_tolerance_simultaneous_change} shows the impact of various modification combinations of all parameters on the dispersion relations (see cation for details). In all panels, the green circles indicate the dispersion curve of the nominal design (set 2) while the black dots show the resulting dispersion relations due to the modification of the parameters. The important observation is that in all configurations, the primary impact is a shift in the DBE frequency (up to $\sim 1\textrm{THz}$) while the shape of the dispersion curve, specifically the flat band near $k\Lambda=\pi$ are retained. Thus, we can conclude that the proposed device exhibits DBE properties that are relatively robust to variations and fabrication errors.

\section{\label{sec:conclusion}Conclusion}
In this paper we presented a comprehensive three dimensional analysis of a periodic structure consisting of two coupled waveguides with grating. We outlined simple guidelines facilitating the optimization of the design parameters for realizing a DBE exceptional point. We use those guidelines to find two sets of parameters exhibiting DBEs, resulting weak and strong coupling between the modes, and compare them. We considered and advantages and drawbacks of the two options, reaching to the conclusion that the strong coupling scheme is advantageous. We calculated the transmission and reflection of a finite structure with parameters corresponding  to strong coupling. We showed that the Q-factor of the DBE resonance of our structure scales as to $N^5$, as opposed to the RBE resonances that scale proportional to $N^3$. This is a clear indication that the formed band-edge is indeed a DBE SP. Finally, we demonstrated the structure’s robustness to geometrical imperfections, showing that the DBE SP in the dispersion relations is preserved and that perturbations are  manifested mainly by shifts in the DBE frequency. The proposed structure can be readily realized using contemporary nano-fabrication facilities with potential applications for high-power optical amplifiers and narrow-linewidth, low-threshold, lasers.
 
\section{Acknowledgment}
The authors thank Prof. Filippo Capolino for stimulating discussions and useful comments. 

This work was partially supported by AFOSR grant no. FA8655-20-1-7052.

\bibliography{bib_2D_gratings}

\end{document}